\documentclass[english]{jpp}
\pdfoutput=1
\usepackage{graphicx}
\usepackage{epstopdf, epsfig}
\usepackage{amsmath}

\shorttitle{ADI preconditioners for the Vlasov equation}
\shortauthor{M. Gasteiger, L. Einkemmer, A. Ostermann, and D. Tskhakaya}

\title{ADI type preconditioners for the steady state inhomogeneous Vlasov equation}

\author{Markus Gasteiger\aff{1}\corresp{\email{markus.gasteiger@uibk.ac.at}}, Lukas Einkemmer\aff{1}, Alexander Ostermann\aff{1}, \and David Tskhakaya\aff{2}}

\affiliation{
\aff{1}University of Innsbruck, Technikerstr.  13, A-6020 Innsbruck, Austria
\aff{2}Technical University of Vienna, Institute of Applied Physics, Fusion@\"OAW, Wiedner Hauptstr. 8-10/E134, A-1040 Wien, Austria
}

\begin{document}
\maketitle

\begin{abstract}
    The purpose of the current work is to find numerical solutions of the steady state inhomogeneous Vlasov equation. This problem has a wide range of applications in the kinetic simulation of non-thermal plasmas. However, the direct application of either time stepping schemes or iterative methods (such as Krylov based methods like GMRES or relexation schemes) is computationally expensive. In the former case the slowest timescale in the system forces us to perform a long time integration while in the latter case a large number of iterations is required.

    In this paper we propose a preconditioner based on an ADI type splitting method. This preconditioner is then combined with both GMRES and Richardson iteration. The resulting numerical schemes scale almost ideally (i.e.~the computational effort is proportional to the number of grid points). Numerical simulations conducted show that this can result in a speedup of close to two orders of magnitude (even for intermediate grid sizes) with respect to the not preconditioned case. In addition, we discuss the characteristics of these numerical methods and show the results for a number of numerical simulations.
\end{abstract}

\section{Introduction}

Particle collisions strongly influence the plasma edge properties in fusion as in technological plasmas.
In magnetic confinement fusion plasma devices the scrape-off layer (SOL) is the plasma region where the field lines are open and end at wall surfaces. The plasma dynamics in this region is markedly different from the confined plasma in the core. This is due to the fact that the plasma is absorbed at the walls. Moreover, in this region there are strong inelastic effects caused by the interaction with neutral and impurity particles (see, for example, \cite{stangeby2000}). In a similar way in technological plasmas collisions strongly affect the quality of the produced materials (see, for example, \cite{liebermann}).
Therefore understanding the inelastic processes in the plasma edge represents one of the hottest topics in plasma research.
This study is especially important for fusion plasma, where the life time of the plasma facing components is defined by plasma edge characteristics (see, for example, \cite{wesson97}).
%Since the plasma wall interaction SOL is the region where the interaction with the walls takes place, understanding the associated plasma and how it strains its surrounding walls is of great importance. This is especially true for the divertor plates, which receive a high portion of the plasma exhaust (see, for example, \cite{wesson97}).

In these edge plasmas the often employed fluid-based models are not sufficient as the assumption of thermal equilibrium does not hold. Thus, a kinetic approach is required. Due to its complexity realistic analytic descriptions of the plasma edge is almost impossible and various numerical tools were developed. These tools represent either particle based codes (see, for example, \cite{tskhakaya2012recent}) or so called Vlasov solvers (see \cite{batishchev1996kinetic}). In general plasma edge modeling kinetic codes are rare and extremely complex, and corresponding simulations demand a lot of computational resources. Therefore the development of new optimized tools for edge plasma modeling is highly important.

In this paper we introduce new optimized solvers for the inhomogeneous Vlasov equation, which includes electron-ion collisions (and is also called the Fokker--Planck equation). The general case, where we have a six-dimensional time-dependent problem (three dimensions in space and three dimensions in velocity), is extremely challenging to compute with current day computer facilities. But in most relevant cases, the system can be reduced to lower dimensions.
In the present paper we continue our previous efforts (see \cite{gasteiger14}) and focus on the $1$+$2$ dimensional steady state model described in section \ref{sec:physical_model}. In the new model we consider more realistic conditions for a much larger plasma system and show effective optimization tools of the Vlasov solver.

Nevertheless, even in the context of this reduced and optimized model numerical solutions incur a significant computational cost. In particular, using a time integration algorithm to determine the steady state solution requires long time integration (the final time has to be proportional to the number of grid points used, as is discussed in section \ref{sec:numerical_methods}). Consequently, we propose a method that uses iterative solvers in order to approximate the steady state. The advantage of these numerical methods is that they can be implemented without ever assembling the corresponding matrices (see, for example, \cite{saad2003}) which significantly reduces their memory footprint. This is particularly important for high-dimensional problems.

Unfortunately, iterative methods can be quite expensive if no good preconditioner is available. In the present paper we propose a way to precondition the inhomogeneous Vlasov equation that is based on an ADI type splitting. This preconditioner yields numerical schemes that scale almost ideally in the number of grid points (i.e.~the run-time is proportional to the number of grid points). These numerical methods are described in section \ref{sec:numerical_methods}. At the end we conduct numerical simulations which show that the ADI type preconditioners outperform the regular algorithms by orders of magnitude.

\section{Physical model\label{sec:physical_model}}

As already mentioned in the introduction we will consider a kinetic model of a quasi-neutral edge plasma. Thus, we start with the inhomogeneous Vlasov equation (for the electrons)
\begin{equation}\label{eq-vlasov-gen}
\partial_t f + \vec{v}\cdot \partial_{\vec{r}}f + \frac{\vec{F}}{m_e}\partial_{\vec{v}}f
= C(f),\end{equation}
%%?? where went the Vlasov Poisson part
%%?? what happened to the Boltzmann eq.
where the electron density $f$ is a function of position $\vec{r}$ and velocity $\vec{v}$, and the electron mass is denoted by $m_e$. The collision term $C(f)$ describes the particle-particle interactions. %This includes electron-electron as well as electron-ion collisions (the ions are assumed to be stationary and thus only enter via the collision term).

In our model of the edge plasma we transform the problem into a coordinate system that is aligned to the magnetic field, which consequently is parallel to the $z$-axis.
This is the same coordinate system that is used in \cite{chodura92}. %%still not verified.
By assuming a homogeneous distribution in the directions perpendicular to the magnetic field we eliminate all but one space coordinate. As a matter of fact, the model itself does not use the magnetic field further. Thus the electric field $\vec{E}$ is the only field introduced into the equations and consequently the electrons experience the force of $\vec{F}=(0,0,e E_z)$, where $e$ is the charge and $E_z$ is the $z$-component of the electric field. Using the quasi-neutrality constraint for simplicity we assume $E_z = \text{const}$. We further assume that the electrons are perfectly magnetized and the corresponding distribution function depends on the parallel ($v_z$)  and total ($v$) velocities. For convenience, below we introduce the cosine $\mu = \cos \theta = v_z/v$ of the pitch angle $\theta$. In this coordinate system the Vlasov equation takes the form %%?? cosine of the angle $\mu ... is an interruption necessary?, eg. an   ,i.e.   ??
\begin{equation} \label{eq:vlasov-newcoo}
    \partial_t f + \mu v \partial_z f + \frac{e E_z}{m_e}\frac{1-\mu^2}{v}\partial_\mu f = C(f).
\end{equation}

%The advantage of the coordinate system being aligned to the field is twofold. First, it makes the symmetry in the problem apparent and allows us to reduce the dimensionality of the problem.
%Second, many collision terms take a simpler form in this coordinate system.
%% comment: should be clear to all readers

Although, for the plasma edge the collision operator in general includes Coulomb (see, for example, \cite{karney1986Fokandquacod}) as well as charged-neutral (for example, we refer to \cite{janev04}) and inelastic charged-charged particle collisions (see, for example, \cite{tskhakaya2015modelling} and the references there), in the present work we restrict ourselves to an electron-ion collision term $C_{ei}$, derived in \cite{rosenbluth57}:
\begin{equation}\label{eq:celecion}C_{ei} = \frac{1}{2}\frac{\gamma_{ei}}{v^3} \partial_\mu\left((1-\mu^2)\partial_\mu f\right), \end{equation}
 where $\gamma_{ei}$ is a constant (in our model) comprising all physically relevant parameters
\begin{equation}\gamma_{ei} = \frac{e^4 n_e L}{4\pi m_e^2 \epsilon_0^2 v_T^3}.\end{equation}
This formula introduces the electron density $n_e$, the Landau logarithm $L$, the vacuum permittivity $\epsilon_0$ and the thermal speed $v_T$.
%This formula requires that we plug in the relevant values for the electron charge $e$, the electron density $n_e$, the Landau factor (basically the minimum deflection angle cutoff), the electron mass $m_e$, the vacuum permittivity $\epsilon_0$, and the thermal velocity $v_T$.
%$\gamma_{ei}$ is a constant during the numerical simulation.
The specific values are listed in table \ref{tab:values}.

For the numerical simulation we render equations (\ref{eq:vlasov-newcoo}) and (\ref{eq:celecion}) dimensionless. Thus, we define the velocity normalized by the thermal speed
\begin{equation}\xi = \frac{v}{v_T}\end{equation}
and measure length as
\begin{equation}\zeta=\frac{z}{Z},\end{equation}
where $Z$ is the length of the spatial domain. These two quantities imply the following normalization of time
\[ \tau = \frac{t}{T}=t\frac{v_T}{Z}. \]
Note that no normalization is required for $\mu$ as this is already a dimensionless variable.

Finally, the dimensionless equation used in the numerical simulation is as follows
\begin{equation}
    \partial_\tau f + \mu \xi \partial_\zeta f + \frac{e E_z Z}{m_e v_T^2}\frac{1-\mu^2}{\xi}\partial_\mu f = \frac{\gamma_{ei} Z}{2 v_T}\frac{1}{\xi^3} \partial_\mu\left((1-\mu^2)\partial_\mu f\right).
\end{equation}
Since we are only concerned with steady state solutions, we drop the time derivative in the present paper and obtain
\begin{equation} \label{eq:final-problem}
    \mu \xi \partial_\zeta f + \frac{e E_z Z}{m_e v_T^2}\frac{1-\mu^2}{\xi}\partial_\mu f = \frac{\gamma_{ei} Z}{2 v_T}\frac{1}{\xi^3} \partial_\mu\left((1-\mu^2)\partial_\mu f\right).
\end{equation}
Combining all the relevant parameters we finally obtain
\begin{equation} \label{eq:motion-dimensionless}
    \mu \xi \partial_\zeta f + \kappa\frac{1-\mu^2}{\xi}\partial_\mu f = \lambda\frac{1}{\xi^3} \partial_\mu\left((1-\mu^2)\partial_\mu f\right),\end{equation}
where we have defined
\begin{equation}\kappa = \frac{e E_z Z}{m_e v_T^2}\end{equation}
and
\begin{equation}\lambda=\gamma_{ei} \frac{Z}{2v_T} = \frac{e^4 n_e L Z}{8\pi m_e^2 \epsilon_0^2 v_T^4}.\end{equation}
Typical values for the dimensionless parameters in edge plasmas can be found in table \ref{tab:values}.

\begin{table}
	\centering
	\hfill
	\begin{tabular}{|c|c|}
		$e$ & $1.6\cdot10^{-19}\text{ C}$\\ % electron charge
		$n$ & $10^{18}\text{ m}^{-3}$\\ % particle density
		$L$ & $10$\\ % Landau coefficient
		$m_e$ & $9.109\cdot10^{-31}\text{ kg}$\\ % electron mass
	\end{tabular}
	\hfill
	\begin{tabular}{|c|c|}
		$\epsilon_0$ & $8.85\cdot10^{-12}\text{ F/m}$\\ % vacuum permias edge plasma in einem Tokamak? Das sollte man	dann auch in der caption der Tabelle sagen.ttivity
		$Z$ & $0.1\text{ m}$\\ % space scaling/system size
		$v_T$ & $2.13\cdot10^{6} \text{ m/s}$\\ % thermal speed
		$E_z$ & $10^3 \text{ V/m}$\\ % electric field
	\end{tabular}
	\hfill
	\begin{tabular}{|c|c|}
		$\gamma_{ei}$ & $1.64\cdot10^6\text{ s}^{-1}$\\
		$\kappa$ & $0.385$\\
		$\lambda$ & $0.0385$\\
	\end{tabular}
	\hfill{}
\caption{All used parameters are shown. They represent typical values for an edge plasma in a fusion device. The derived coefficients are also listed.\label{tab:values}}
\end{table}

From a physical point of view equation (\ref{eq:motion-dimensionless}) has three contributions. The first term on the left hand side describes an advection along the magnetic field lines with speed $v_z/v_T$. The acceleration due to the static electric field is modeled by the second term on the left hand side. Notice that this results in an advection for $\mu$ as the electric field along the $z$-axis increases/decreases the velocity along $z$ and thus increases/decreases $\mu$. Finally, the right hand side describes the diffusive effects due to the (almost) stationary ions scattering the electrons.

\begin{figure}
	\centering
	\includegraphics[width=\textwidth]{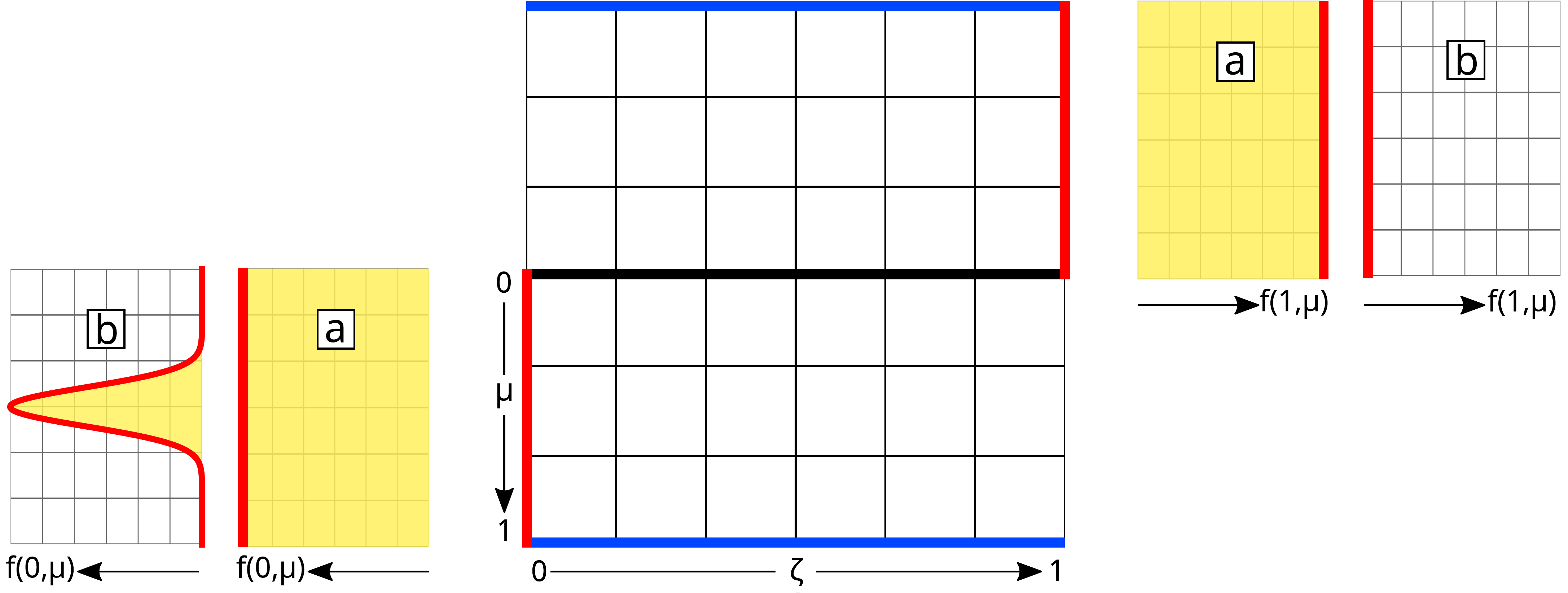}
    \caption{\label{fig:boundaries}In this illustration the 2D grid in $\zeta$ and $\mu$ is shown with the imposed boundary conditions. Boundaries with homogeneous Neumann boundary conditions are colored blue, whereas those with Dirichlet boundary conditions are colored red. Uncolored boundaries are outflow boundaries. Next to the inflow boundaries two possible boundary data are sketched. These isotropic and electron beam inflow conditions are tagged with \fbox{a} and \fbox{b}, respectively.}
\end{figure}

To set the boundary condition for the above problem we have to keep in mind that we want to describe a realistic domain for edge plasmas, i.e.~plasma as it can be observed e.g. in the SOL. Due to the assumed symmetry in the angle like coordinate $\mu$ we infer that any electron flux crossing the $\mu$ boundary must be balanced by a flux from the other side of the boundary. Thus, we will use homogeneous Neumann boundary conditions in this case. For the $z$-boundary we have to make a distinction between the left boundary (i.e.~the boundary at $\zeta=0$) where the inflow is determined by the relevant plasma dynamics in the interior of the device. The actual value is highly problem dependent, but we generally use Dirichlet boundary conditions there. On the other hand, for the right boundary (i.e.~the boundary at $\zeta=1$) we have to model the wall of the device. As a first approximation we use homogeneous Dirichlet boundary conditions there. However, more involved boundary conditions that reflect a certain amount of plasma are certainly a viable (and probably more realistic) alternative that we will explore in the future. The boundary configurations used in the following examples are illustrated in figure \ref{fig:boundaries}.
% \newlength{\ab}
% \settoheight{\ab}{ab}
% \newcommand*{\mystrut}{\rule{0pt}{.9em}}

\section{Numerical methods\label{sec:numerical_methods}}
\renewcommand{\vec}[1]{#1}

In this section we will describe the numerical methods used to obtain an efficient approximation to equation (\ref{eq:motion-dimensionless}). To that end we will first describe the discretization and then discuss the iterative methods that are used to solve the resulting linear system. The latter is particularly important in order to obtain an efficient numerical method.

Let us note that $f$ depends on $\zeta$, $\mu$ and $\xi$. However in the present case $\xi$ is a parameter only and we thus have to solve a two dimensional problem in $\zeta$ and $\mu$ repeatedly (once for each value of $\xi$). It should be kept in mind, however, that depending on the value of $\xi$ the relative strength of diffusion and advection can differ significantly.

First, the diffusive term is discretized by the following finite difference approximation
\begin{equation}
\label{eq_diff_stencil}\partial_\mu\left((1-\mu^2)\partial_\mu f\right)\approx \frac{1}{h_\mu^2}\left((1-\mu_{i+1/2}^2)f_{i+1} - (2-\mu_{i+1/2}^2-\mu_{i-1/2}^2)f_i+(1-\mu_{i-1/2}^2)f_{i-1}\right).
\end{equation}
In the last equation we denoted the equidistant $\mu$-step size by $h_\mu$ and $f_i = f(\zeta,\mu_i)$, where $\mu_i= i  h_\mu$ for $i\in \{-n_\mu,\ldots,n_\mu\}$ with $ n_\mu h_\mu = 1$.

In principle, any second order accurate stencil can be used here. Note, however, that the stencil above results in a symmetric matrix. As has been pointed out in the previous section, Neumann boundary conditions are used for both the boundary at $\mu=1$ and the boundary at $\mu=-1$. The additional $\mu$-advection term, which is the second term on the left hand side of equation (\ref{eq:motion-dimensionless}), is discretized with centered finite differences. Note, however, that this term degenerates and thus no boundary conditions are required.

In contrast, for the $\zeta$-advection we use an upwind scheme (i.e.~a different stencil is employed for positive and negative $\mu$, respectively). This facilitates the implementation of the Dirichlet boundary conditions and guarantees that we have perfect outflow conditions in the regions where this is important (we refer again to figure \ref{fig:boundaries}).

The resulting linear system can be solved with an arbitrary numerical method. However, our goal is to develop an efficient approach. Although we only consider the $1$+$2$ dimensional case in the present paper, it is an important consideration that such schemes can be generalized to the higher dimensional setting. In this context it is of particular interest to discuss numerical methods that scale well with respect to memory consumption. Consequently, we will employ so-called iterative methods, which can be implemented matrix-free. That is, only the application of the matrix to a vector has to be computed (and there is no need to assemble and store the entire matrix in memory). The disadvantage of these methods, however, is that they usually require a large number of iterations which results in a slow algorithm. In many applications this problem can be overcome by employing a suitable preconditioner.

In the described linear problem we have a matrix $A$ and a right hand side $\vec{b}$. In the present case $\vec{b}$ has very few non-zero elements, which are due to the boundary terms.
Our goal is then to solve
\begin{equation}\label{key_a}
    A \vec{x} = \vec{b}
\end{equation}
by some iterative method (see, for example, with \cite{saad2003}).

A preconditioner $M$ is a matrix that can be easily inverted and constitutes an approximation to $A$. For a right preconditioner we set $ \widetilde{A}=A M^{-1}$, $y = M x$ and consider the linear system
\begin{equation}
\widetilde{A} \vec{y} = \vec{b},
\end{equation}
which has to be solved for $\vec{y}$. The solution of this preconditioned system hopefully requires less iterations as ideally $A M^{-1}$ is close to the identity.  The unknown $\vec{x}$ can then be determined easily by solving $M\vec{x}=\vec{y}$.

Similar to this approach a left preconditioner takes the form
\begin{equation}\label{left_preconditioner}
    M^{-1} A \vec{x} = M^{-1} \vec{b}.
\end{equation}
This, however, has the disadvantage that the residual is modified which causes some complications in deciding when to stop the iteration. Consequently, we will only use right preconditioners in the present work.

%In the examples we used 2 different $M$, which are described in section \ref{sec:numerical_results}.

Let us now turn our attention to constructing an efficient preconditioner. For elliptic and parabolic problems preconditioners based on multigrid methods (\cite{briggs2000}) or domain decomposition (\cite{smith2004}) can be used. While some of these methods have been generalized to hyperbolic problems (in particular, in the context of fluid models; see, for example, \cite{Adams2010,Shadid2010}) significant challenges remain. Consequently, in recent years so-called physics based preconditioners have become popular (see, for example, \cite{Chacon2003,chacon2008,reynolds2010,reynolds2012}). These preconditioners exploit the particular structure of the equation under consideration and are often successful where more generic algorithms are inefficient or fail altogether.

In our problem, the matrix $A$ can be partitioned into terms that only contain derivatives in the $\mu$ and $\zeta$ directions, respectively. Thus, we can write our matrix as $A = A_{\zeta} + A_\mu$. This partitioning is the basis of all the preconditioners developed in this paper. The linear system corresponding to the matrices $A_{\zeta}$ and $A_\mu$ can be solved efficiently (using, for example, the Thomas algorithm) as these are only one-dimensional problems. Put differently, we consider preconditioners based on dimension splitting.

As a first step, one might argue that the diffusion found in $A_\mu$ is what limits the convergence speed. Thus, we define $M=A_\mu$. In the following we refer to this as the $\mu$-preconditioner. While this approach results in some improvement compared to the iterative schemes without any preconditioner the scaling is still far from optimal.

Therefore, we will consider a second preconditioner based on an alternating direction implicit (ADI) approach (such methods have a long tradition in the time integration of partial differential equations; see, for example, \cite{yanenko1971,lindemuth73,schnack80} or for more recent work \cite{hujeirat1998,sgura2012,reynolds2012,muller2014}). We define
\begin{equation}\label{key_b}
    M = (1-s A_\mu) (1 - s A_{\zeta} ),
\end{equation}
where $s$ is a tunable parameter. Note that this is equivalent to a Lie--Trotter splitting step in a time integration scheme (with step size $s$). Since performing a numerical time integration yields a successively better approximation to the steady solution, we can use this splitting scheme as a preconditioner. This approach combines the advantages of iterative methods (which force a relaxation to the equilibrium of interest) with a time marching scheme (which allows us to construct an efficient preconditioner).

At first sight we might be tempted to choose a large $s$ (since an approximation at a later time yields a better approximation to the steady state solution). However, due to the splitting error committed there is a trade-off to be made here. In the numerical simulations conducted we found that values on the order of $0.1$ give the fastest convergence.

Both of these preconditioners can be implemented without storing the corresponding matrix and suitable $\mathcal{O}(n)$ algorithms are available for each step. Thus, the overhead in computational cost is roughly a factor of three compared to the unpreconditioned case. As we will see in section \ref{sec:numerical_results} the ADI preconditioner reduces the number of iterations by two orders of magnitude. Consequently the algorithm based on the ADI preconditioner is significantly less computationally demanding.

Both of these preconditioners are generic enough that they can be combined with any iterative numerical method. In particular, we will consider Richardson iteration and the generalized minimal residual method (GMRES) (see, for example, \cite{saad2003}). The latter is part of the class of Krylov subspace methods and usually converges faster than relaxation schemes. This performance comes at the cost that we have to store multiple vectors of previous iterations in memory. In contrast, Richardson iteration can be implemented using a single input and a single output array and is thus extremely memory efficient.

To conclude this section, let us discuss an approach to find the steady state solution that is commonly used in the literature. Namely, to integrate the time dependent equations of motion (\ref{eq:motion-dimensionless}) sufficiently long in order to obtain an approximation to the steady state. However, in the present case this approach suffers from severe difficulties. This is due to the fact that the present model includes an advection in the $\mu$ direction the speed of which can be arbitrarily small (in the numerical discretization the smallest speed is proportional to the grid size $h$). Since we have to resolve the corresponding timescale, the final time has to scale as $1/h$ (which implies that the final time of the integration is large). On the other hand, performing splitting is attractive as it allows us to obtain an efficient time integrator without a CFL constraint. Moreover, such methods have been studied extensively (see, for example, \cite{cheng1976,klimas1994,qiu2011,rossmanith2011,filbet2001,arber2002,crouseilles2010,einkemmer2014convergence,einkemmer2014,casas2016,crouseilles2015hamiltonian,crouseilles2016asymptotic,einkemmer2016structure}). However, the error made by the time integrator then dictates a relatively small time step (due to accuracy requirements) which implies that we have to take a large number of steps in order to integrate to the final time. Consequently, such methods are not computationally efficient for the problems considered here. Note that, in particular, the GMRES approach gives us more freedom in this regard as, in a manner of speaking, we can take different time steps for different parts of the problem.

\section{Numerical results\label{sec:numerical_results}}

\begin{figure}
	\includegraphics[width=0.48\textwidth]{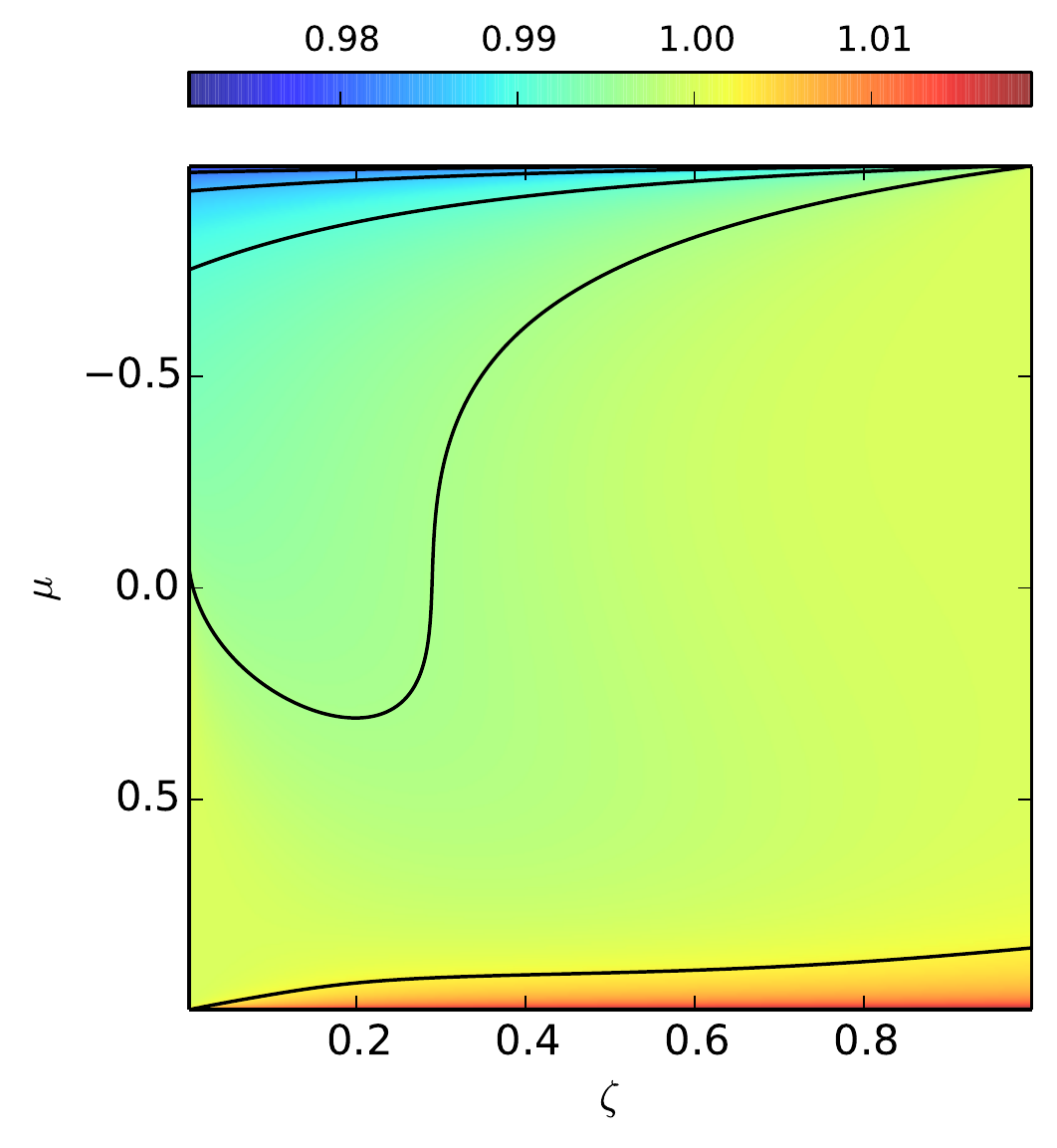}
	\hfill
	\includegraphics[width=0.48\textwidth]{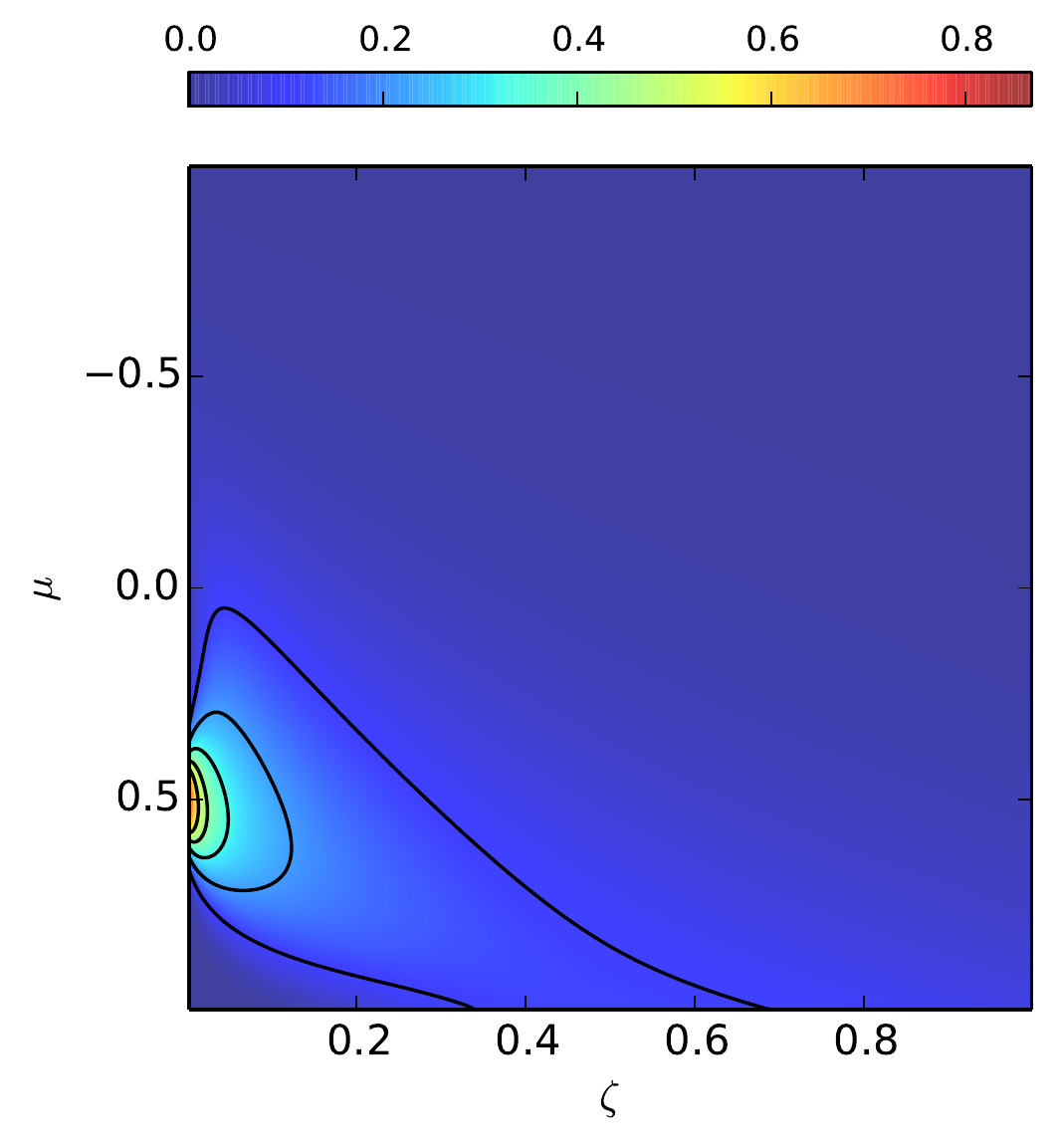}
	\caption{On the left we see the numerically computed steady state when using isotropic Dirichlet boundary condition for the inflow at $\zeta=0,1$. The results on the right are obtained by using an electron beam instead. To create these plots we used $\kappa=0.385$, $\lambda =0.0385$, $\xi=0.5$.\label{result}
	}
\end{figure}

To compare the various numerical methods we compute a solution to equation (\ref{eq:motion-dimensionless}) for different grid sizes. We also consider two different inflow boundary conditions. In particular, we show results for the following two inflow boundary conditions
\begin{enumerate}
\item \textit{isotropic particle distribution.} The corresponding Dirichlet boundary conditions take the constant value $1$ as is illustrated in figure \ref{fig:boundaries}.
\item \textit{electron beam.} In this case we model an electron beam that enters the edge plasma but is not aligned to the magnetic field lines. For the corresponding Dirichlet boundary conditions a Gaussian beam profile given by
        \[ f(\mu,\zeta)=
	        \begin{cases}
		        0, & \mu < 0, \zeta=1,\\
		        \mathrm{e}^{-100\cdot (\mu - 0.5)^2}, & \mu \geq 0, \zeta=0 \\
	        \end{cases}\]
        is prescribed.
\end{enumerate}
Numerical results for both of these boundary data are shown in figure \ref{result}.

\begin{figure}
	\centerline{\includegraphics[width=\textwidth]{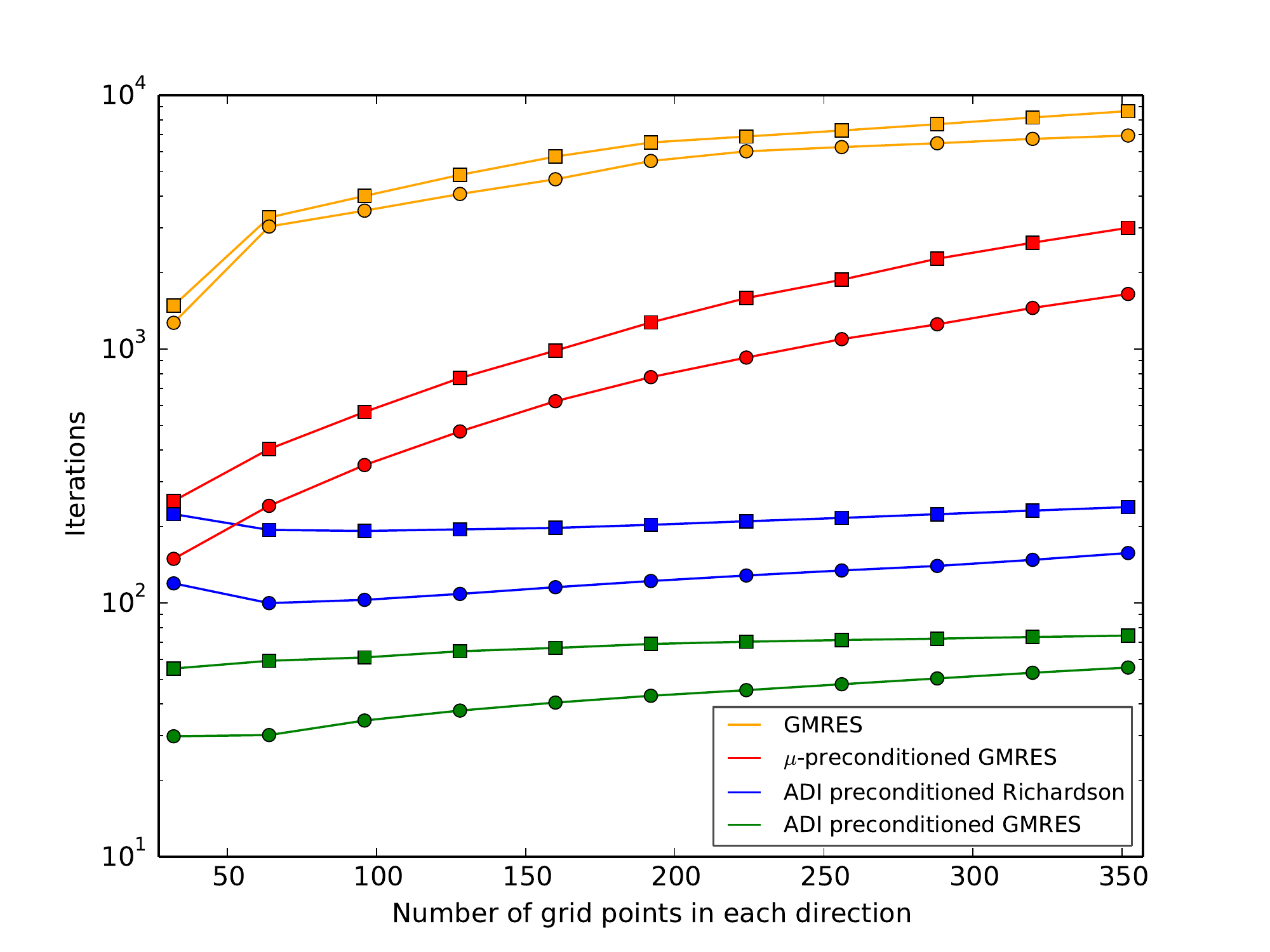}}% Images in 100% size
    \caption{The number of iterations each numerical method requires to reach a precision below $10^{-5}$. These numbers are obtained by averaging the results of the parameters $\kappa=0.385$, $\lambda =0.0385$, and $\xi=\{0.125, 0.25, 0.5, 0.75, 1.0, 1.25, 1.5, 2.0, 3.0, 4.0, 5.0, 6.0\}$. The parameter in the ADI scheme was set to $s=0.1$.
    The $\square$ and $\circ$ respectively mark the results obtained with isotropic and beam initial values, as described in the text.}
	\label{fig:iterations}
\end{figure}

All algorithms stop when a certain precision is reached (in all the numerical simulations considered here we require that the residual is below $10^{-5}$). We only show results with an equal number of grid points in both the $\mu$ and $\zeta$ direction. Thus, we can compare the number of iterations each method requires to achieve a given precision (averaged over a range of values for $\xi$). The corresponding results are plotted in figure \ref{fig:iterations}.

We can also observe from figure \ref{fig:iterations} that the number of iterations required differs slightly with respect to the boundary condition used.
We conclude that all preconditioners proposed in this paper improve the rate of convergence significantly. In addition, we find that
\begin{enumerate}
    \item Only inverting $A_\mu$ is the cheapest preconditioner but also the least effective one. While this preconditioner might be competitive for small grids, as we increase the number of grids points it scales as $N^{2.2}$, where $N$ is the number of grid points in one direction.
    \item The ADI preconditioner works very well and scales almost ideally as we increase the number of grid points. As a result this preconditioner is able to reduce the number of iterations required by two orders of magnitude (compared to the unpreconditioned GMRES implementation). This is true independent of the iterative method used (i.e.~both GMRES as well as Richardson iteration show this behavior). However, as expected, the GMRES implementation significantly outperforms the implementation based on Richardson iteration (by approximately a factor of $3$) (see \cite{saad2003} for instance). This comes at the cost of increased memory consumption (in the simulation we employ a restarted GMRES scheme with Krylov dimension $20$).
\end{enumerate}

\section{Conclusions}

We proposed a method for the solution of the $1$+$2$ dimensional steady state inhomogeneous Vlasov equation. This method has been based on an iterative numerical method which is preconditioned using an ADI type splitting scheme. We show that this preconditioners succeed in reducing the number of iterations required (and thus the execution time) by orders of magnitude. We also find that the resulting iteration scheme scales almost ideally as we increase the number of grid points.

The new preconditioning method allows one to consider large size plasma edge systems in a realizable timescale and can be used for kinetic modeling of realistic edge plasma systems.

As future work we will extend the methods introduced in this paper in two directions. First, we will consider physical models in higher dimension (i.e.~go beyond the $1$+$2$ dimensional case considered here). This makes the use of high performance computing infrastructure essential (in this context we should emphasize that the methods described here lend themselves very well to parallelization). Second, we will consider a wide range of commonly used collision terms (both for electron-electron as well as for electron-ion collisions) with the goal of finding preconditioners that can be efficiently applied and result in a significant reduction of the computational effort.

\newpage

\section{Acknowledgment}
%The first author is supported by the \emph{Friedrich Schiedel Foundation for Energy Technology}.
M. Gasteiger is a fellow of the Friedrich Schiedel Foundation for Energy Technology.

%\section{References}
\bibliographystyle{jpp}
\bibliography{jpp}

\end{document}